\documentclass[11pt, conference]{csce}

\usepackage[hmargin=.8in,vmargin=1in]{geometry}
\usepackage[american]{babel}
\usepackage[T1]{fontenc}
\usepackage{times}
\usepackage{caption}
\usepackage{textcomp}
\usepackage{epsfig,graphicx}
\usepackage{xcolor}
\usepackage{amsfonts,amsmath,amssymb}
\usepackage{fixltx2e} 
\usepackage{booktabs}
\usepackage{subfiles}
\usepackage{enumerate}
\usepackage{url}
\usepackage{hyperref}
\graphicspath{{figures/}}

\usepackage{fancyhdr}

\urldef{\mailsa}\path|kamyar.nazeri@uoit.ca|
\title{\bf Effect of Social Media on Opinion Formation}

\author{
	{\bfseries Kamyar Nazeri}\\
	Faculty of Science, University of Ontario Institute of Technology\\
	\mailsa
}

\begin{document}

\maketitle

\begin{abstract}
In this work, we investigate the effect of social media on the process of opinion formation in a human population. This effect is modeled as an external field in the dynamics of the two-dimensional Sznajd model \cite{sznajd2000opinion} with a probability $p$ for an agent to follow the social media. We investigate the evolution of magnetization $m(r)$, the distribution of decision time $\mathbb{P}(\mu)$ and the average relaxation time $\langle \tau \rangle $ in the presence of the external field. Our results suggest that the average relaxation time on the lattice of size $L$ follows a power law form $\tau \sim L^{\alpha}$, where the exponent $\alpha$ depends on the probability $p$. We also found that phase transition between two distinct states of the system decreases for any initial distribution of the opinions as the probability $p$ is increasing. For a critical point of $p \sim 0.18$ no phase transition is observed and the system evolves to a dictatorship regardless of the initial distribution of the opinions in the population.
\end{abstract}

\vspace{1em}
\noindent\textbf{Keywords:}
 {\small Sociophysics, Opinion Dynamics, Sznajd Model, Ising Model, Social Media \\}

\section{Introduction}

	Social physics or Sociophysics have been studied for many applications in the past twenty years. It was developed by theoretical physicists who mainly came from statistical physics to understand the complex behavior of human crowds like how opinions are formed. These models are based on two simple hypothesis: 1) in any social system, laws on the microscopic scale can explain phenomena on the macroscopic scale. In this scheme, models that describe the microscopic behavior of the system are used to infer the general large-scale behavior of the system and 2) in many cases, individuals in a society behave like a particle which have no feelings and no free will \cite{cialdini2009influence}. That means particle or agent based models can be used to simulate the behavior of individuals and the law of large number suggests that by performing the simulation a large number of times the results are close to the expected value. Opinion dynamics is a class of these agent-based models concerning social opinion changes through local interactions among individuals. To simulate opinion dynamics, two models are often proposed in the literature based on social physics: (i) \textit{the Voter model} \cite{liggett2012interacting} is based on continuous time Markov process of an interacting particle system. In this model each person selects the opinion of the 	neighborhood, with a probability proportional to the number of \textit{n} neighbors having that opinion \cite{stauffer2009opinion}; and, (ii) \textit{the Snzajd model} \cite{sznajd2000opinion} which is based on a simple psychological behavior of the people in a closed community: The fundamental way that an individual decides what to do in a situation is to look to what others are doing. This phenomenon is known as \textit{social validation} and is proposed in a model called ``United we Stand, Divided we Fall'' or USDF \cite{sznajd2000opinion}. In contrast to the voter model where the probabilities depend linearly on \textit{n}, here they depend exponentially on \textit{n}. Sznajd model uses Ising spin system ($\pm 1$) to simulate opinions of agents.
	
	In \cite{sznajd2005sznajd} Snzajd-Weron introduced a modification to the original model and showed how the same model could be applied in politics, marketing, and finance. Increasing the number of variable's states in the Sznajd model and the range of interactions have also been explored in the literature \cite{stauffer2002sznajd,deffuant2000mixing}. In \cite{sznajd2005left} Snzajd-Weron extended her original idea to personal versus economic opinions by assigning two Ising spins to each site (agent) and they were able to divide people into four clusters: Socialist, Libertarians, Authoritarians, and, Conservatives. The effect of the mass media, modeled as an external stimulation acting on the social network on the process of opinion formation is investigated in \cite{crokidakis2012effects} and \cite{grabowski2006ising}, in their work they modeled the process of opinion formation as a scale-free network with agents of different degrees, taking into account their spatial localization.
	
	In this work, we simulate opinion dynamics in a closed community using a 2D Sznajd model on the lattice of size $L \times L$. We investigate the effect of social media on the process of opinion formation in the human population. This effect is modeled by imposing as an external field in the dynamics of the Sznajd model with a probability $p$ for an agent to follow the social media. We consider a situation where all agents are exposed uniformly to the social media effect and they would follow the social media with a probability $p$; we will further show how to model a system with a non-uniform social media exposure by introducing a conditional probability $P_S$ that agents share contents on a social network and have influence on their neighbors opinions.

\section{One-Dimensional Sznajd Model}

\begin{figure}[!htb]
		\centering
		\includegraphics[width=1\linewidth]{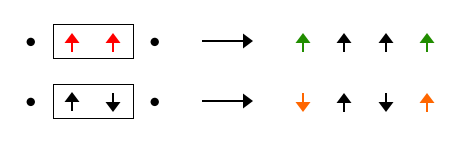}
		\caption{Schematic representation of one-dimensional Sznajd model. Only spins inside the rectangle participate to the dynamics. }
		\label{fig:1d}
	\end{figure}

\begin{figure}[!htb]
		\centering
		\includegraphics[width=.9\linewidth]{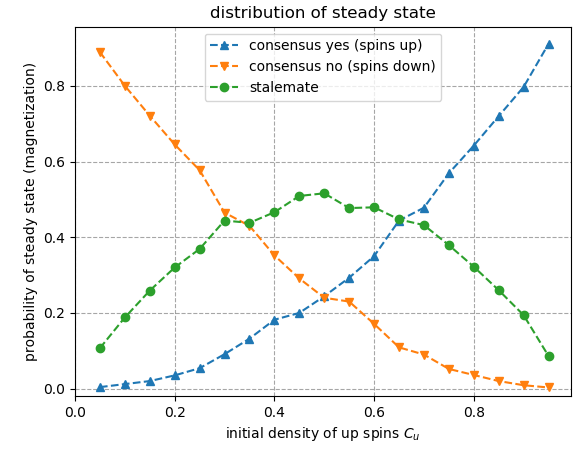}
		\caption{Distribution of the steady-state (magnetization) as a function of initial density of up spins $C_u$. Averaging was done over 1000 samples. }
		\label{fig:1d_steady}
	\end{figure}

\begin{figure}[!htb]
		\centering
		\includegraphics[width=.9\linewidth]{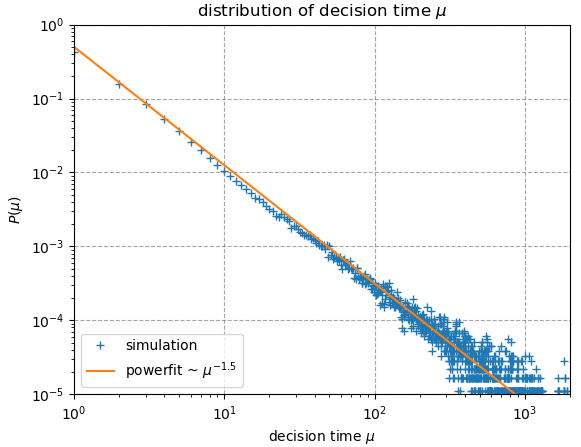}
		\caption{Distribution of decision time $\mathbb{P}(\mu)$ follows a power law with an exponent $\sim -\frac{3}{2}$. Averaging was done over 1000 samples. }
		\label{fig:1d_decision_time}
	\end{figure}

\begin{figure}[!htb]
		\centering
		\includegraphics[width=1\linewidth]{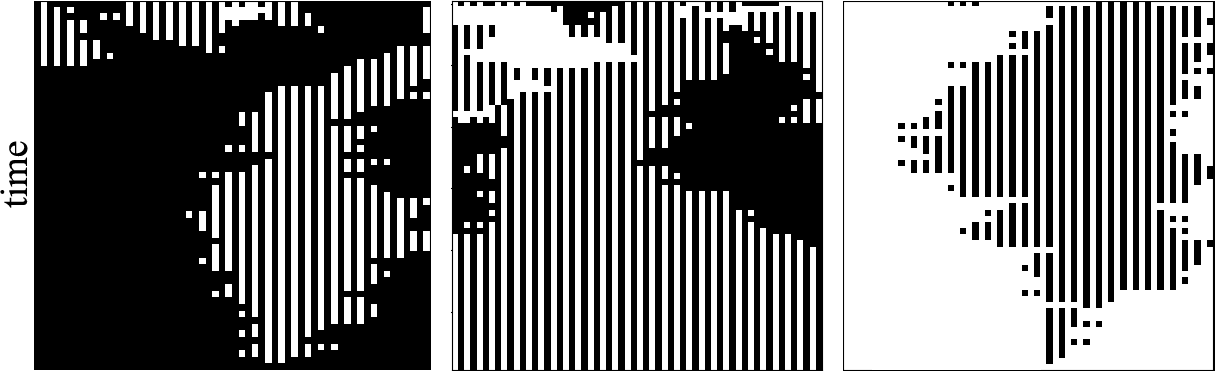}
		\caption{The time evolution binary heat-map of the model states with random initialization. consensus no (left), stalemate (center), consensus yes (right)  }
		\label{fig:1d_heatmap}
	\end{figure}

\begin{figure}[!htb]
	\centering
		\includegraphics[width=.9\linewidth]{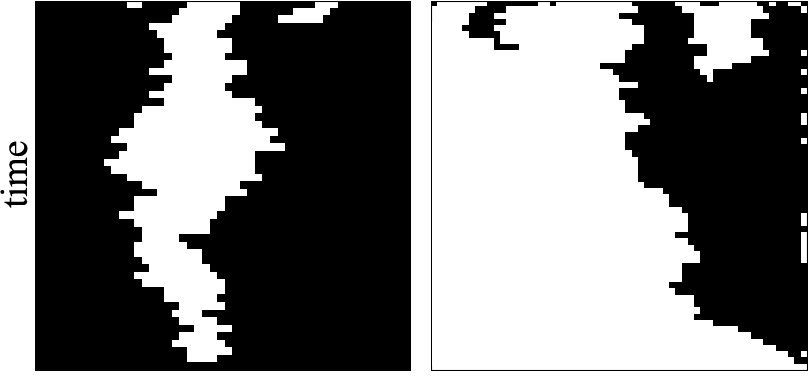}
		\caption{The time evolution binary heat-map of the model states for a modified Sznajd model. consensus no (left), consensus yes (right)  }
		\label{fig:1d_heatmap_modified}
	\end{figure}

\begin{figure}[!htb]
		\centering
		\includegraphics[width=1\linewidth]{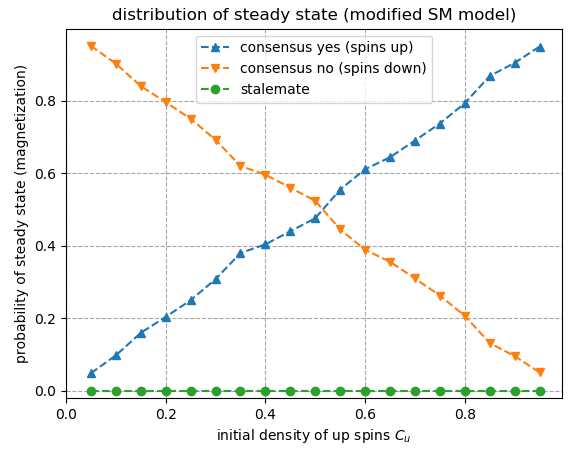}
		\caption{Magnetization versus initial density of up spins $C_u$ for a modified Sznajd model. Averaging was done over 1000 samples. }
		\label{fig:1d_steady_modified}
	\end{figure}

	The Sznajd model is based on Ising spin  (`yes' or `no') system describing a mechanism of making up decisions in a closed community \cite{sanchez2004modified}. The dynamic rules of the Sznajd model are \cite{sznajd2000opinion}: 
	\begin{enumerate}[i)]
		\item In each step a pair of spins $S_i$ and $S_{i+1}$ is randomly chosen to change their nearest neighbors.  i.e. the spins $S_{i−1}$ and $S_{i+2}$.
		\item If $S_i$ = $S_{i+1}$ then $S_{i-1}$ = $S_i$ and $S_{i+2}$ = $S_i$.
		\item If $S_i$ = $-S_{i+1}$ then $S_{i-1}$ = $S_{i+1}$ and $S_{i+1}$= $S_i$.
	\end{enumerate}
	(See the schematic representation of Fig. \ref{fig:1d})
	\newline
	We consider the one-dimensional Sznajd model as our baseline and compare our findings against it. To investigate the model, standard Monte Carlo simulations with random updating is performed. In an isolated community, with uniform distribution of initial states (opinions), there are only two possible steady-states possible: complete consensus or stalemate. Depending on the initial concentration of up spins $C_u$ the system can reach ferromagnetic state (complete consensus) or antiferromagnetic (stalemate) with different probabilities Fig. \ref{fig:1d_steady}. It is worth mentioning that with 50\% initial concentration of up spins (`yes'), the system reaches `stalemate', `all yes' or `all no' states with a probability of 50\%, 25\% and 25\% respectively. Notice that to obtain a steady-state of `all yes' with probability of 50\%, initial concentration of $C_u \gtrsim 0.7$ is required.
	\vspace{-10px}
	It was shown in \cite{sznajd2000opinion} that by having these dynamic rules in the system, the distribution of the decision time $\mathbb{P}(\mu)$ follows a power law with  an exponent $\sim -\frac{3}{2}$  (Fig. \ref{fig:1d_decision_time}). In other words, if an individual changes his (her) opinion at time $t$, he (she) will probably change it also at $t+1$, that's why $\mu$ is often very short. However, sometimes individuals stay for a long time without changing their opinions. 
	\vspace{-10px}
	\newline
	The stalemate (antiferromagnetic) state of the one-dimensional Sznajd model is not realistic for a human society. To achieve exactly a 50-50 final state in a community is
	almost impossible, specially if it is composed by more than a few dozens of	members \cite{sanchez2004modified}. The time evolution binary heat-map of the Snzajd model is shown in Fig. \ref{fig:1d_heatmap}, note that even with the cases of complete consensus, the unrealistic behavior of antiferromagnetism (striped regions) is rampant as the system evolves in time.
	\\
	To avoid this unrealistic behavior, new dynamic rules were proposed in \cite{sousa2006outward} and later incorporated into Sznajd model in \cite{sznajd2005sznajd}:
	\begin{enumerate}[i)]
		\item In each step a pair of spins $S_i$ and $S_{i+1}$ is randomly chosen to change their nearest neighbors.
		\item If $S_i$ = $S_{i+1}$ then $S_{i-1}$ = $S_i$ and $S_{i+2}$ = $S_i$.
		\item If $S_i$ = $-S_{i+1}$ then $S_{i-1}$ = $S_{i}$ and $S_{i+2}$= $S_{i+1}$.
	\end{enumerate}
	\noindent \newline
	Using the new rules, in case of disagreement of the pair $S_i$  and $S_{i+1}$ , the spin \textit{i} ends up with at least one neighbor having its own opinion. It can be seen in Fig. \ref{fig:1d_heatmap_modified} that antiferromagnetism almost never happens toward reaching the steady-state in the system.
	The magnetization process (steady-state) for a modified Sznajd model is shown in Fig. \ref{fig:1d_steady_modified}. As expected stalemate is not found in the simulation and the system always reaches full consensus regardless of the initial distribution of opinions. It can be seen that the relationship between the magnetization and the initial distribution is almost linear, with a phase transition happening at $C_u = 0.5$. ie. to obtain a steady-state of `all yes' with probability of 80\%, initial concentration of $C_u \gtrsim 0.8$ is required.

\section{Two-Dimensional Sznajd Model}

\begin{figure}[!htb]
		\centering
		\includegraphics[width=1\linewidth]{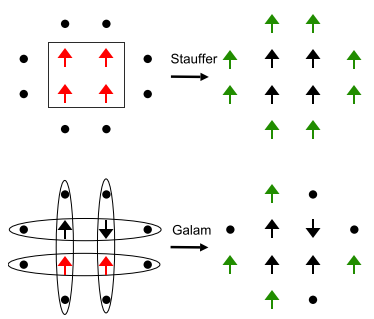}
		\caption{Schematic representation of two-dimensional Sznajd model proposed by Stauffer (top) and Galam (bottom). A panel of $2 \times 2$ spins are randomly selected and influence neighboring spins denoted by dots. }
		\label{fig:2d}
	\end{figure}

\begin{figure}[!htb]
		\centering
		\includegraphics[width=1\linewidth]{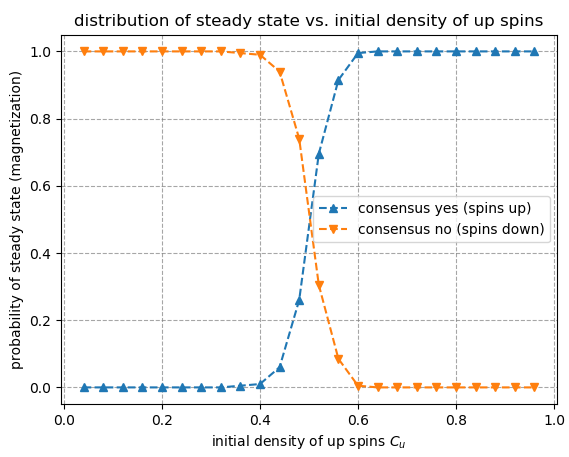}
		\caption{Magnetization versus initial density of up spins $C_u$ for a two dimensional Sznajd model on a square lattice. Averaging was done over 1000 samples. }
		\label{fig:2d_steady}
	\end{figure}

	Two dimensional models are more useful and realistic for social systems; to analyze the effect of social media on a closed community and a better approximation for social interactions, two-dimensional Sznajd models on a square lattice of the size $L \times L$ are proposed \cite{stauffer2000generalization,galam2008sociophysics} where again every spin can be up or down with random updating rules. Galam \cite{galam2008sociophysics} showed that the same updating rule of the one-dimensional Sznajd model can be applied in the two-dimensions in the following way: at each step a 2x2 panel of four neighbors is randomly selected and the one-dimensional rule is applied to each of the four pairs to change their neighbors opinions (see Fig. \ref{fig:2d}). In \cite{stauffer2000generalization} Stauffer proposed a method with the same outcome which is computationally more efficient:
	\begin{enumerate}[i)]
		\item A 2x2 panel of four neighbors is randomly selected.
		\item The panel persuades its eight neighbors to follow the orientation if and only if all spins are parallel.
	\end{enumerate}
	Stauffer dynamic rules of two-dimensional Sznajd model prevents system from reaching stalemate. For the initial concentration of the up spins  $C_u > 0.5$ (< 0.5) the system goes into a ferromagnetic state with all spins up (down) which characterizes a phase transition at $C_u \approx 0.5$ (see Fig. \ref{fig:2d_steady}). 
	This phase transition separates two distinct states of the system: for $C_u < 0.5$ the model never reaches the full consensus with all spins up while with $C_u > 0.5$ full consensus is always reached.

\begin{figure}[!htb]
		\centering
		\includegraphics[width=1\linewidth]{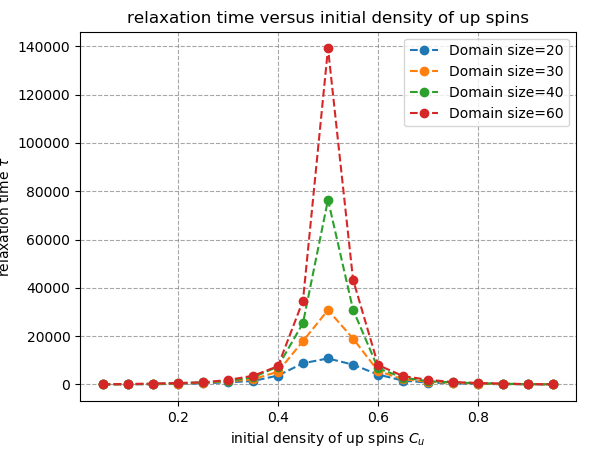}
		\caption{Relaxation time $\tau$ versus initial density of up spins $C_u$ for a two dimensional Sznajd model on a square lattice. Averaging was done over 1000 samples. }
		\label{fig:2d_relaxation_time}
	\end{figure}

\begin{figure}[!htb]
		\centering
		\includegraphics[width=1\linewidth]{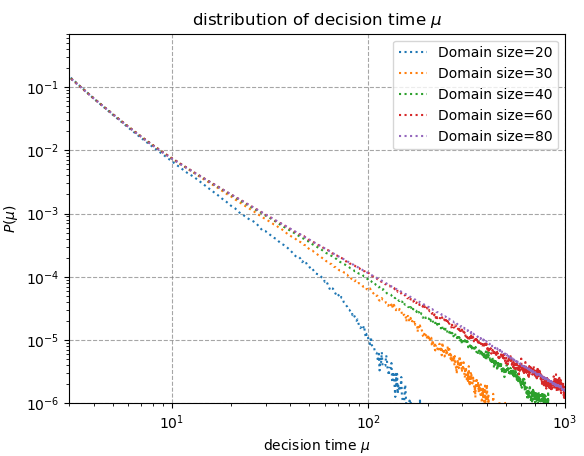}
		\caption{Distribution of decision time $\mu$ in a two dimensional Sznajd model on a square lattice for different values of domain size. Averaging was done over 500 samples. }
		\label{fig:2d_decision_time_domain}
	\end{figure}

	The time needed to reach a complete consensus in this model is called \textit{relaxation time} and is denoted by $\tau$. It is interesting to see how relaxation time changes in a two-dimensional Sznajd for different values of $C_u$ (see Fig. \ref{fig:2d_relaxation_time}). As expected at the phase transition point $C_u = 0.5$ the simulation takes the longest to reach to a steady-state, we'll show later that the relationship between the relaxation-time $\tau$ and $C_u$ follows a power law. Note that for small (or large) values of the $C_u$, the system reaches full consensus relatively fast. This suggests that in the absence of the external force the system remains at equilibrium state for $C_u \approx 0.5$ and it takes a long time to reach to a steady-state.
	\newline
	Unlike the one-dimensional Sznajd model the distribution of the decision time for two dimensional model does not always follow a power-law (see Fig. \ref{fig:2d_decision_time_domain}). The short time distribution of decision time follows a power law with exponent $\approx −1.8$ whereas the longtime distribution of the decision time is exponential and depends on the domain size. The effect of domain size in a two-dimensional model is like an information noise shown in the one-dimensional model \cite{sznajd2000opinion}. This suggests that the domain size needs to be large enough $ L \gtrapprox 60 $ in a two-dimensional system, in order to minimize the effect of noise in the simulation. For large domain sizes, the distribution of the decision time is a good fit to the line $\tau^{-1.8}$ in the log-log plot.
	
\begin{figure}[!htb]
		\centering
		\includegraphics[width=1\linewidth]{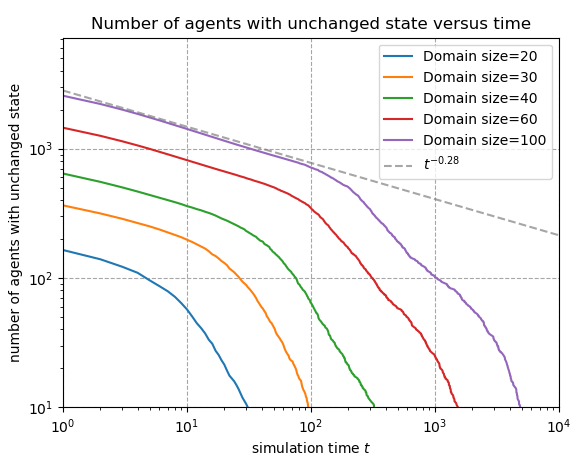}
		\caption{ Number of never changed agents for different values of domain size. Averaging was done over 500 samples. }
		\label{fig:2d_unchanged_domain}
	\end{figure}

	\noindent
	In Fig. \ref{fig:2d_unchanged_domain} the number of never changed spins $N_{uc}$ versus time $t$ is presented. In this case a power law relaxation with exponent of $\omega \sim -0.28$ is found for time $t < t^*$ where $t^*$ depends on the domain size $L$. For values of $t > t^*$, $N_{uc}$ exponentially decays to zero. \\\\
	$
	N_{uc} = 
	\begin{cases} 
	t^{-\omega} & \text{for } t < t^*, \omega \sim 0.28 \\
	\exp(\alpha t) & \text{for } t > t^* \\
	\end{cases}
	$
	\noindent \newline
	As expected, number of never changed agents decays slower for larger values of the domain size.

\section{Social Media Simulation}

\begin{figure}[!htb]
		\centering
		\includegraphics[width=.85\linewidth]{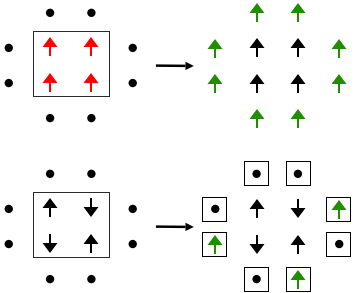}
		\caption{Schematic representation of two-dimensional Sznajd model with an external field: A panel of $2 \times 2$ spins are randomly selected and influence neighboring spins denoted by dots (top) or each agent independently of the others is influenced by the social media. (bottom)}
		\label{fig:2d_social}
	\end{figure}

	We consider two-dimensional Sznajd model on a square lattice of size $L$ and periodic boundary condition with $N=L^2$ agents and Ising spins of $\pm1$ for each site. We model the effect of social media as an external field applied on each site uniformly. The following rules govern our model:
	\begin{enumerate}[i)]
		\item At each step, a $2\times 2$ panel is randomly selected.
		\item If all four center spins are parallel, the eight nearest neighbors are persuaded to follow the panel orientation. 
		\item If not all four center spins are parallel, consider the influence of social media (external field): each one of the eight neighbors follows, independently of the others, with probability $p$.`
	\end{enumerate}
	\noindent\newline
	The effect of social media is considered toward +1 opinion; in other words, at each time step each neighbor will change his (her) opinion to +1 with a probability of $p$ if he (she) could not be persuaded by the $2 \times 2$ agents. Note that in our model, the effect of social validation (group of agents) is stronger than the effect of the social media i.e. an individual is more likely to follow friends, relatives or colleagues and is only influenced by the social media if there's no consensus among them. (see Fig. \ref{fig:2d_social})
	\subsection{Visualization of Opinion Formation}
\begin{figure}[!htb]
		\centering
		\includegraphics[width=1\linewidth]{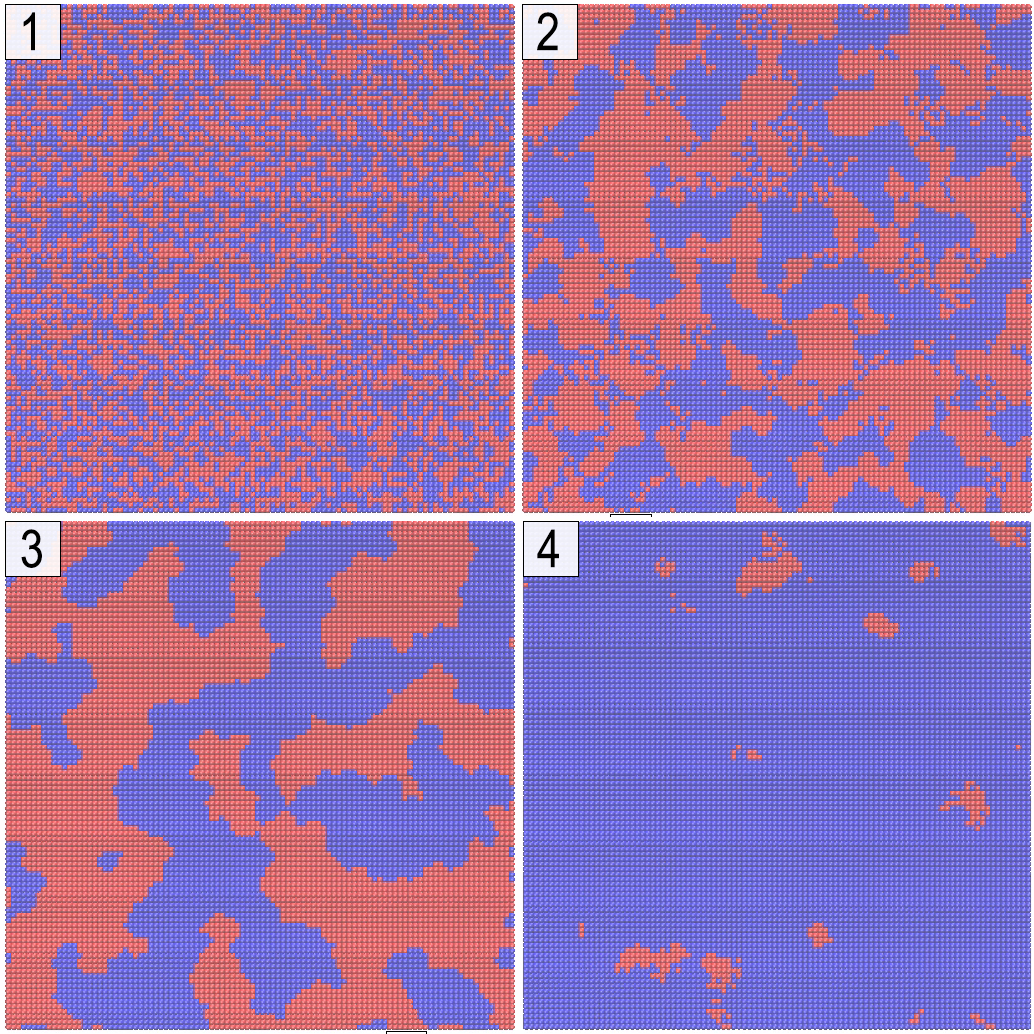}
		\caption{ Dynamic evolution of two-dimensional Sznajd model on a lattice of $L=100$ with initial concentration of up spins $C_u = 0.5$ and external force $p=0.1$ at different time-steps: $t = 0$ (1), $t = 100$ (2), $t = 1000$ (3), $t = 2000$ (4) where one time-step is equal to $10^4$ Monte Carlo steps. Blue and Red agents represent `yes' (1) and `no' (-1) opinions respectively. Visualization is performed in Visual Molecular Dynamics (VMD) software package.}
		\label{fig:vmd}
	\end{figure}

	For simplicity, we assume that all agents are exposed uniformly to the social media. We can visualize the effect of social media on the social opinions of the system to see how opinions are formed (see Fig. \ref{fig:vmd}). Here the distribution of the opinions is visualized at different time-steps for lattice size $L=100$ and initial concentration os up spins $C_u=0.5$. The external field is applied to the agents with a probability of $p=0.1$. The $-1$ and $+1$ states of agents are visualized with `red' and `blue' colors respectively. It can be seen that ``\textit{United we Stand Divided we Fall}'' nature of Sznajd model causes agents with the same opinion to form clusters to survive. At $t=2000$ system is almost reaching full consensus. Note that the simulation is reaching consensus almost 10 times faster in presence of the external field ($p=0.1$), we'll show later in this work how $p$ relates to the relaxation time $\tau$.
	
	\subsection{Social Media Effect on Magnetization}
	Following te standard Sznajd model we study the magnetization process with respect to the initial concentration os up spins in presence of an external field:
	$$
	m(r) = \frac{1}{N} \sum_{i=1}^{N} S_i
	$$	
	Where $N=L^2$ is the total number of agents in the simulations and $S_i=\pm1$.
	\begin{figure}[!htb]
		\centering
		\includegraphics[width=1\linewidth]{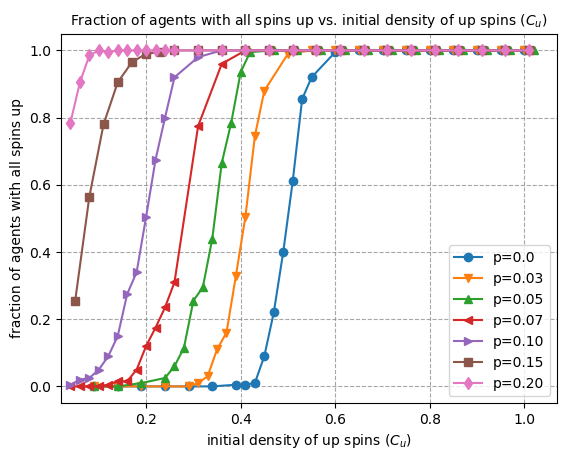}
		\caption{Fraction of agents with all spins up versus the initial density of up spins $C_u$ for different probabilities $p$ of social media influence (domain size $L=50$). Averaging was done over 500 samples.}
		\label{fig:2d_steady_social}
	\end{figure}
	\noindent \newline
	The effect of social media (external field) on the magnetization is shown in Fig. \ref{fig:2d_steady_social}. Here, the fraction of agents with all spins up are shown as a function of initial concentration os up spins $C_u$ which represent the probability of the system reaching the full consensus. We are explicitly interested in the phase transition observed in the standard two-dimensional Sznajd model. In other words, given a certain value of the probability $p$ we want to find the $C_u$ at which the outcome of the system is changing. Note that $p=0$ represents the standard two-dimensional Sznajd model. We have considered 500 samples for each probability and took the average. \\	
	It can be seen from Fig. \ref{fig:2d_steady_social} that increasing the probability $p$, progressively decreases the phase transition point. With some probability $p_{thresh}$ that the system does not present a phase transition. e.g. for $p=0.2$ regardless of the initial concentration of up spins $C_u$ the system always reaches the full consensus with all spins up. Our simulations suggests that with a domain size of $L \gtrsim 50$ phase transition only happens for $p \lesssim 0.18$. i.e. to win in an election regardless of what each individual's opinion is at the beginning, a party only needs to influence mass society with a probability of $p=0.18$ and winning will be almost certain. \\
	
	\subsection{Social Media Effect on Relaxation Time}
\begin{figure}[!htb]
		\centering
		\includegraphics[width=.95\linewidth]{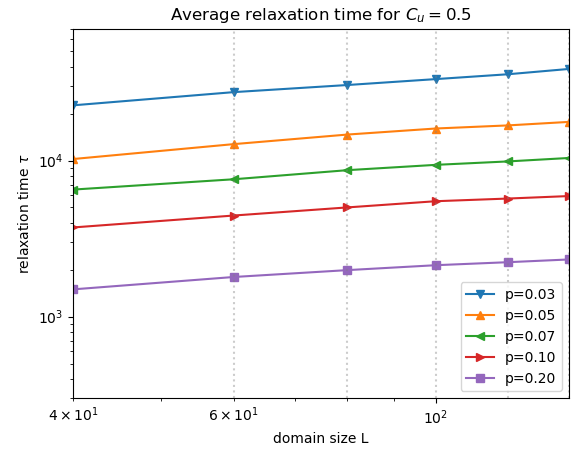}
		\caption{ Average relaxation time $\tau$ versus domain size $L$ with initial concentration of up spins $C_u = 0.5$ for different probabilities $p$ of social media influence. Averaging was done over 500 samples.}
		\label{fig:2d_relaxation_time_social}
	\end{figure}

	To compare the simulation results against the standard two-dimensional Sznajd model, the relaxation time $\tau$ is measured for different values of $p$ (see Fig. \ref{fig:2d_relaxation_time_social}). as shown in \cite{sznajd2000opinion} in a one-dimensional Sznajd model the relationship between the relaxation time and the domain size $L$ in a two-dimension model is log-normally distributed as observed in other versions of Snzajd model \cite{stauffer2000generalization,crokidakis2012effects,crokidakis2010consequence}. 
	\noindent \newline
	The results of relaxation time $\tau$ versus domain size $L$ is illustrated in Fig. \ref{fig:2d_relaxation_time_social} for different values of $p$ in the log-log scale. It can be seen that the relationship between the relaxation time $\tau$ and the domain size follows a power law form $\tau \sim L^{\alpha}$ where $\alpha$ depends on the probability $p$. Note that $p=0$ is the standard two-dimensional Sznajd model, whereas increasing $p$ causes a decrease in the average relaxation time in the log scale.
	\begin{figure}[!htb]
		\centering
		\includegraphics[width=.85\linewidth]{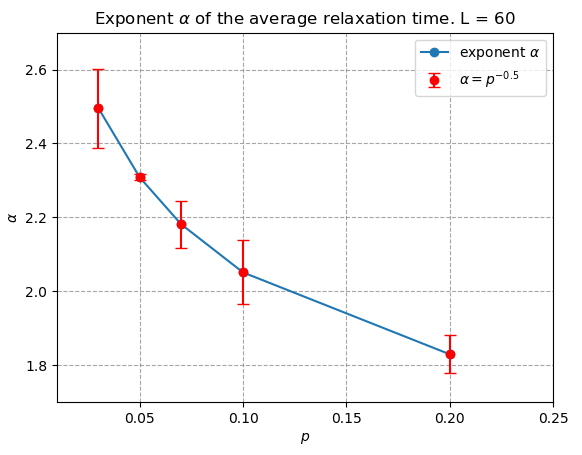}
		\caption{ Power exponent $\alpha$ for different probabilities $p$ on a lattice size of $L=60$. $\alpha$ decreases by increasing $p$ and the relationship follows a power law. Averaging was done over 500 samples.}
		\label{fig:2d_exponent_social}
	\end{figure}
	It seems interesting to find a relationship between the exponent $\alpha$ and probability $p$. Multiple regression analysis on the average relaxation time exponent $\alpha$ for different values of $p$ on a lattice of size $L=60$ shows that this relationship is also a power law with the form $\alpha \sim p^{-0.5}$ (see Fig. \ref{fig:2d_exponent_social})
	\noindent \newline
	
	\subsection{Social Media Effect on Decision Time}
\begin{figure}[!htb]
		\centering
		\includegraphics[width=.9\linewidth]{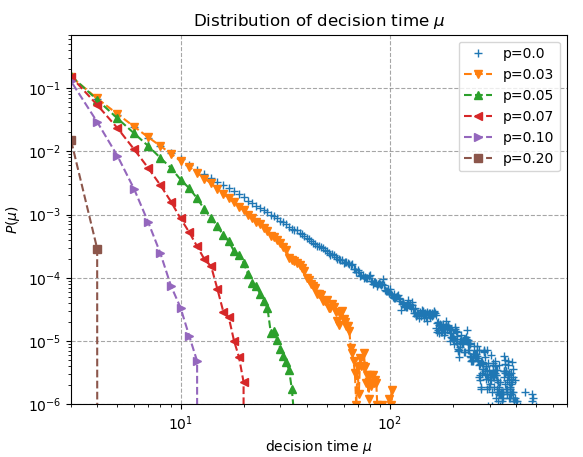}
		\caption{ Distribution of decision time $\mathbb{P}(\mu)$ for different values of $p$: Power-law for small values of $\mu$ and exponential for large values of $\mu$. Averaging was done over 500 samples.}
		\label{fig:2d_decision_time_social}
	\end{figure}

	As illustrated before (Fig. \ref{fig:1d_steady}) the distribution of the decision time $\mathbb{P}(\mu)$ follows a power law for a one-dimensional Sznajd model \cite{sznajd2000opinion,sznajd2005sznajd}; However, the same pattern cannot be generalized to the two-dimensional model in the presence of an external field (see Fig \ref{fig:2d_decision_time_social}). 
	
	Our simulations suggests that there is some characteristic time $\mu^*$ which depends on a value of $p$. For decision times less than $\mu^*$ the distribution is a power-law of the form $\mu^{-1.8}$ and for values of $\mu > \mu^*$ it follows an exponential form $\exp(\alpha\mu)$ where $\alpha$ depends on the probability $p$. These results suggests that the effect of an external field is similar to adding noise (similar to ``social temperature'') as presented in \cite{sznajd2000opinion} for a one-dimensional Sznajd model. Reaching full consensus faster means more individual are being exposed to the external field and consequently more people are changing their opinions. 
	
	\begin{figure}[!htb]
		\centering
		\includegraphics[width=1\linewidth]{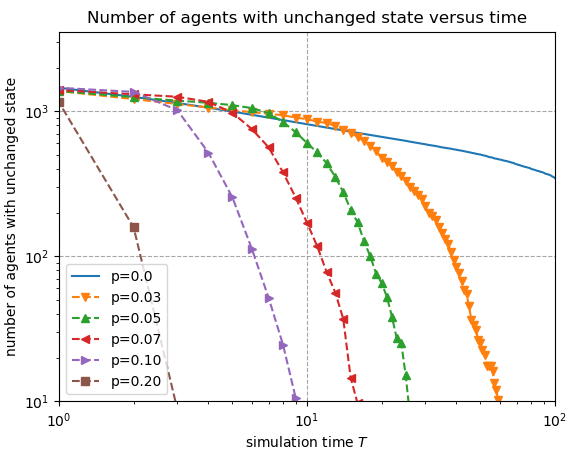}
		\caption{ Number of agents with unchanged states over time for different values of $p$. Averaging was done over 500 samples.}
		\label{fig:2d_unchanged}
	\end{figure}
	\noindent 
	The number of agents with unchanged state $N_{uc}$ versus time is presented in Fig. \ref{fig:2d_unchanged}, it can be seen that this distribution follows the same pattern as before: there is a transition point in the distribution of ``number of never changed agents'' and it depends on the system size as a power law. It suggests that there's a direct relationship between the number of agents changing states and the decision time where $N_{uc}$ changes behavior from power law to exponential at some characteristic time $t^*$. Similar results were presented in the standard two-dimensional Sznajd model (see Fig. \ref{fig:2d_decision_time_domain}): an agent can stay for a long time without changing his mind, however, if he changes his opinion at time $t$ he will probably change it also at time $t+1$ \cite{sznajd2000opinion}. Once enough agents change state, the change in final state of the system becomes inevitable: one change can cause an avalanche and the social media facilitate this change.

\section{Non-Uniform External Field}

	So far we have assumed that every agent is exposed uniformly to the social media; this scenario is not very realistic in an actual human society. In a human society, individual agents linked together to form a complex mesh of interconnected relationships, and while each agent gets affected by its neighbors through social validation (the information flows inward), it can influence its neighbors by exposing them to social media. e.g. in a social network, people might expose their friends by sharing the news that is inclined to have a particular opinion. We can extend our model by introducing a probability $P(S)$ that an agent share news on social media ($P(S)$ is considered a probability toward +1 opinion). Any agent only reacts to the external field with probability $P(F)$ if it is already exposed to it. This model can be formulated by a conditional probability $P(F|E)$ where condition $E$ is the probability that an agent is already exposed to some news. $P(E)$ can also be formulated by a conditional probability $ P_{\text{neighbor}}(S|E) $. We can relax this conditional probability by assuming that the two events are IID; the probability of getting affected by the social media then becomes:
	\begin{equation}
		P_{sm} = P(F|E) = P(F)P(E) \propto P(F)P(S)
	\end{equation}
	It means that probability of getting affected by the social media $p_{sm}$ is proportional to the product of probabilities of sharing news $P(S)$ and reacting to an external field $P(F)$ and the proportionality is measured by the initial exposure of the community to the external field. \\
	It can be seen from (1) that sharing news on the social network directly affects $p_{sm}$ and we showed that $p_{sm} \gtrsim 0.18$ is needed to change the outcome of the system regardless of the initial concentration of the opinions. That means sharing news (fake news) on the social network, even when the end goal is to criticize the news or give negative feedbacks, directly contributes to $p_{sm}$ and one step ahead toward full consensus. In fact, the best action to take when exposed to a ``fake news'' is to ignore it!

\newpage
\section{Conclusion}

	In this work, we studied the effect of social media on opinion formations. We took the Ising-model-based approach of opinion dynamics and developed a modified version of the two-dimensional Sznajd model. We modeled influence of social media by introducing a probability $p$ that an agent would follow the social media in the dynamics of the standard Sznajd model. In our model, all agents are exposed to the external field uniformly and $p=0$ recovers the standard Snzajd model; Our simulations suggest that phase transition in magnetization happens only for $p \lesssim 0.18$. i.e. an influence probability of $p \gtrsim 0.18$ is needed to win a two-party election regardless of the initial density of the opinions among members of the society. We further showed that we can model more complex systems of opinion formations by introducing a probability that an individual is exposed to the social media. \\\\
	We also studied the effect of external force on the relaxation time $\tau$, as shown by \cite{sznajd2000opinion} in a one-dimensional Sznajd model the relationship between the relaxation time and the domain size $L$ in a two-dimension model is log-normally distributed. In our simulations, we found that the average relaxation time on the lattice of size $L$ follows a power law form $\tau \sim L^{\alpha}$ where $\alpha$ also depends on the probability $p$. Using regression analysis we found that the exponent $\alpha$ depends on $p$ and the relationship is also a power law with the form $\alpha \sim p^{-0.5}$. \\\\
	Finally, we found that the distribution of the decision time $\mathbb{P}(\mu)$ in the presence of an external field no longer follows a power law as described in one-dimensional Sznajd model \cite{sznajd2000opinion,sznajd2005sznajd}. Our simulations suggest that this distribution consists of two parts: for decision times less than $\mu^*$ the distribution is a power-law of the form $\mu^{-1.8}$ and for values of $\mu > \mu^*$ it follows an exponential form $\exp(\alpha\mu)$ where $\alpha$ depends on the probability $p$. These results suggest that the effect of an external field is similar to adding noise (similar to ``social temperature'') as presented in \cite{sznajd2000opinion} for a one-dimensional Sznajd model. 
	\\\\\\

\addcontentsline{toc}{section}{references}
\bibliography{references}{}
\bibliographystyle{IEEEtran}

\end{document}